# Lessons Learnt from Expert-Centred Studies Exploring Opportunities and Challenges for Immersive Forensic Investigation

Vahid Pooryousef[*]  Tim Dwyer[†]  Richard Bassed[‡]  Maxime Cordeil[§]
Monash University  Monash University  Monash University  The University of Queensland
                                      Victorian Institute of
                                      Forensic Medicine

Lonni Besançon[¶]
Linköping University

## ABSTRACT

Research studies involving human participants present challenges, including strict ethical considerations, participant recruitment, costs, and many human factors. While human-computer interaction researchers are familiar with these challenges and current solutions, expert-centred studies can be even more challenging in ways that researchers may not anticipate. This issue is particularly important as research grants are increasingly based on practical and real-world problems, which necessitate close collaboration with experts. In this paper, we reflect on and discuss the challenges, solutions, and specific requirements that arose during our expert-centred studies conducted over three years of a PhD study exploring immersive forensic investigation.

## 1 INTRODUCTION

Human-centred research is essential in exploring opportunities for emerging technologies in application domains, to ensure the safety, validity, and practicality of proposed solutions. However, it is challenging because of strict ethical considerations, participant availability and recruitment, costs, as well as many other human factors such as diversity, interest, and experience in the subject.

Researchers working in the field of human–computer interaction typically follow a relatively straightforward process: they design a human-centred study (typically a user study), submit ethics applications, amend and obtain the approval within a reasonable time frame, and then proceed to conduct workshops, interviews, focus groups, and user studies with participants. This often yields a sense of certainty and confidence regarding the requirements, necessary timeframes, and methodological approaches required for conducting human-centred studies.

Nevertheless, with the growing emphasis on practical and applied research, particularly in the last few years, driven by the imperative to secure competitive grants and funding to support research endeavours, it has become increasingly important to identify and address the challenges and potential solutions of collaborating with subject matter experts and stakeholders across specific domains, as well as with external departments and partner organisations. This is especially prevalent in studies where factors such as data confidentiality and sensitivity, the willingness or reluctance of particular departments and individuals to participate, differing research ethics processes, and the research setting or environment, become significant considerations. These factors are particularly important when collaborating with partner institutions.

As a result of these factors, the process of preparing for and conducting expert-centred studies is likely more complex than studies involving non-experts from a wider public. If anything, it presents very specific challenges that are not likely to arise with non-expert participants. In this paper, we will reflect on and discuss our practical experience gained from an industry-funded PhD study centred on the theme of Immersive Forensic Investigation.[1] We aim to provide guidance for future researchers based on our experiences and reflections, thus enabling them to better anticipate issues related to study timing, unforeseen obstacles, effective budget management, and potential solutions that may arise when engaging in research with external partners and stakeholders.

## 2 CONTEXT

We studied the potential applications and development of immersive technologies for end-users in forensic autopsy examination and crime scene investigation. To this end, we collaborated with a forensic medicine institution and a police department.

To investigate the use of immersive technologies in forensic autopsy, we conducted the iterative, participatory design of an immersive visualisation platform for autopsy settings [19]. A forensic autopsy room is a contaminated environment, which makes it cumbersome to use computing systems; however, XR headsets, which allow touchless interaction, provide a significant advantage. While it was useful, it was less likely that XR headsets would be used during an autopsy solely for visualisation and access to imaging data. Thus, we complemented this by facilitating the documentation and report generation process [16], which increases the value of using this technology. All these design and evaluation processes involved the medico-legal experts in an iterative process, including interviews, a workshop, and user studies.

We also explored the use of XR technologies for crime scene reconstruction. Previous studies have investigated the use of XR to view crime scenes in 3D space; however, these were mostly static. Therefore, we developed an XR toolkit for reconstructing animated crime scenes that provides a spatio-temporal context [17]. In this project, we visited and collaborated with police personnel to understand the state of the art in forensic investigation within an advanced forensic services department and to build prototypes informed by expert insights through a co-design process.

## 3 REFLECTION AND DISCUSSION

In this section, we reflect and discuss the challenges and solutions that we faced during the aforementioned studies. We focus on the challenges that we believe do not normally exist in regular human-centred studies with non-expert samples. Where they are common in both, we emphasise the differences.

---

[*]e-mail: vahid.pooryousef@monash.edu
[†]e-mail: tim.dwyer@monash.edu
[‡]e-mail: richard.bassed@vifm.org
[§]e-mail: m.cordeil@uq.edu.au
[¶]e-mail: lonni.besancon@gmail.com

[1]This research is described in the following papers: [18, 16, 17].

### 3.1 Ethics Approval

Scientific research is frequently required to be accomplished within a limited, and often strict, time window due to the typically short timeframes allocated by grants budgets. As a result, obtaining research ethics approval becomes one of the major challenges and often highly time-consuming in expert-centred studies, particularly in cases where the research is conducted in collaboration with departments or organisations outside the traditional academic environment and involves particular stakeholder groups. This added complexity is especially apparent when such external departments do not have established processes or the necessary experience in dealing with ethics approvals.

In our research, for example, even though the relevant forensic medicine institute had entered into an official collaboration agreement with our research team, and the studies we proposed were not especially sensitive in nature, the process of securing approval from this particular institute exceeded four months in duration. This is in contrast to the typical timeframe required for obtaining similar approval at a university, which would usually be completed in approximately two weeks. Nevertheless, we recognise that some universities may have significantly slower processing times due to logistical factors, though this is uncommon among the top universities with which we have had collaboration. There are several contributing factors that may account for this marked difference in approval times, but two primary reasons stand out as particularly significant.

The first major contributing factor may relate to the limited agility of the target institute's committees. Since the forensic medicine institute is not fundamentally a research-focused institution, unlike a university, it is less accustomed to reviewing and granting approval for human research studies, nor does it possess the necessary administrative or technological infrastructure to enable a streamlined, expedited review process. The second factor likely has to do with the unfamiliarity of the subject matter, specifically, immersive technologies, on the part of the personnel at the target institute. In contrast, universities tend to host a range of experts and specialists who already have sufficient knowledge and experience concerning emerging research topics. Accordingly, the marked discrepancy in approval timelines observed between the ethics committees of different types of institutions should be considered by researchers during the planning and implementation of research studies.

Due to the extended and unpredictable length of this process, we were also compelled to avoid making the ethics application excessively detailed, choosing instead to maintain sufficient flexibility for potential adjustments during the developmental phase of our study. Such flexibility is typically less problematic when rapid approval can be secured for low-risk studies. In our other experience, working with the police department was even more challenging compared to the forensic medicine institution. Conducting a user study at the university involving police personnel required approval from the police ethics committee and not the university. However, since we did not have an official collaboration with the police department, obtaining this approval was not easy and, in our time frame, possible. The flexibility approach as described before, remains subject to ongoing debate, especially in the context of clinical trials, since an overly flexible or non-specific ethics application could potentially be exploited for improper purposes. In clinical or high-risk research, therefore, such flexible applications are not recommended as it may be misused (see e.g., the case in clinical trials [8, 15]).

### 3.2 The Ethics of Authorship

In our cases, experts have contributed a lot of knowledge and time to our work and studies, to a degree that could perhaps be seen as being enough for co-authorship. When considering the ethical aspects of authorship in research, we propose that various academic and professional domains should extend authorship opportunities to subject-matter specialists. These experts bring unique perspectives and knowledge that can significantly enhance the quality of research in their respective fields.

Our primary argument highlights the particular difficulties encountered by researchers seeking ethics approval, especially for projects involving sensitive topics or collaboration with external institutions. Such challenges can impede both research progress and quality. However, by formally involving such specialists as authors, researchers can foster closer, more ethical partnerships that are mutually beneficial. This approach helps to ensure adherence to ethical standards and also contributes to a more nuanced understanding in the research design process. Moreover, offering authorship to specialists encourages deeper engagement and accountability, resulting in more robust and comprehensive research findings. This strategy has been successfully implemented in prior studies, leading to more insightful and beneficial outcomes for all stakeholders involved, as evidenced in the literature [23].

However, there are other aspects to consider with respect to the issue of authorship. First, expert-centred studies that Human-Computer Interaction (HCI) and visualisation researchers often conduct imply that the involved experts have to evaluate the prototypes or systems being developed. In this case, there could arise some conflict of interest for them to positively evaluate the system to increase the chances of the research manuscript being accepted. One workaround for this problem is to only offer co-authorship after the systems or prototypes have been evaluated. In other words, HCI and VIS researchers can plan to involve the experts as co-authors from the beginning, but only reveal that after all evaluations have been conducted. It was done, for instance, by Besançon et al. [3, 4] in their work involving surgeons who all became co-authors after their evaluation was conducted. Another challenge is to properly highlight and credit the contributions of these experts in the manuscript. This issue fits right in the ongoing authorship VS contributorship debate [20] where the CRediT author statement[2] is often used to specify the responsibilities and contributions of each of the authors. However, in its current form, the CRediT author statement does not really offer a space for expert users to specify their contributions and may have to be amended to include it. Furthermore, while the CRediT statement is rarely used in visualisation and HCI research venues, it is being encouraged by some (e.g., JoVI [6]).

### 3.3 Availability

One significant issue, frequently encountered when working with small groups of experts (or generally special groups), concerns not only the limited number of interested participants but also their varying levels of availability and flexibility in scheduling. The challenges extend further to the practical problem of accessing these experts in order to conduct the study. Such obstacles are not unique to this research; rather, they have been widely reported in previous studies within the field of HCI that involve expert participants [7, 5, 14, 22, 23]. As such, it is important to explicitly acknowledge that small sample sizes are to be anticipated in these types of studies. This recognition is a well-established norm in the field and should be regarded as an acceptable methodological consideration. By making this clear, researchers can effectively address any potential concerns regarding the rigour or validity of the study that readers may have due to the limited number of participants.

Furthermore, these challenges tend to become even more pronounced in studies employing a cyclical or iterative design, such as those involving multiple rounds (e.g., [19, 16]). In such studies, it is not only the initial difficulty of recruiting a sufficient number of expert participants that presents a challenge. The problem is compounded by the necessity of re-engaging these same individuals for subsequent phases, which requires additional willingness and availability on their part. Experts often have limited spare time and may find it difficult to commit to participation over an extended period or across several rounds of research. This dual challenge of having low participant numbers and inconsistent participant engagement across iterative stages is an issue in this form of research that should always be carefully considered. Where alternative methods are likely to be effective, undertaking such studies should be avoided.

---

[2]https://www.elsevier.com/researcher/author/policies-and-guidelines/credit-author-statement

In addition, studies that require the physical presence of participants significantly reduce accessibility. Therefore, the requirement for physical attendance should only be imposed when deemed necessary. In our second study [16], during the second phase, we utilised videoconferencing to improve accessibility and to compensate for the absence of some first-phase participants. This approach was also chosen because in-person studies tend to overly focus on the usability and limitations of XR technology, which is a recurring theme in many studies and was also examined in our first phase. However, the aim of this second phase was specifically to concentrate on the features and capabilities of the technology, and it was an exploratory study. As a result, while software and hardware glitches are not encountered in this context and the authentic sensation of using the technology is not induced (as experienced in our other studies), greater accessibility, the inclusion of participants from other states or countries, and an understanding of the potential applicability of the technology in various locations are achieved, despite the inability to reassess some usability aspects.

### 3.4 Diversity

In the context of evaluating HCI systems, it is highly common to collect information such as gender, age, and user experience. In our context, user experience with mixed reality technologies is particularly important. However, when conducting evaluations or even designing with specialists in small groups, there are considerable constraints on participant diversity with respect to these factors, to the extent that imposing any meaningful limitations or controls may not be feasible, which has been seen in other participant groups in HCI as well [12]. As a result, study outcomes are often narrowly bound to the specific group and their particular experiences. This introduces a significant risk of bias in interpreting the results and notably limits the generalisability of the findings.

Although there is no straightforward solution for this issue, it is important to acknowledge the limitation explicitly. Furthermore, such investigations should be treated as exploratory in nature. In addition, for emerging technologies, it may be beneficial to place greater emphasis on exploring use cases or features where user experience has a minimal influence on performance. For instance, in [16], we employed video-based evaluation methods. This approach allowed us to minimise challenges associated with direct technological interaction, such as tapping a virtual button in mid-air, thereby reducing the impact of unfamiliarity with the technology on performance. Moreover, this strategy enabled remote participation, further increasing the flexibility of recruitment.

### 3.5 Human Factors

Throughout this project, and in collaboration with forensic specialists, we identified many contributing human factors during the design, development, and evaluation of our immersive forensic investigation tools. In the following, we highlight some of the main factors:

**Understanding potentials through high-fidelity prototypes.** In co-design or participatory methods, low-fidelity techniques, such as paper prototyping and sketching, have become common. However, in our experience—particularly in the first study we conducted [19]—it became clear that these techniques are not always practical, especially when designing for immersive technologies. Many participants, including potential end-users, are not very familiar with these technologies and find it difficult to conceptualise spatial interactions using simple sketches, or understand the capabilities of these technologies. This disconnect often leads to confusion, reduced engagement, and shallow feedback and ideation, as participants struggle to bridge the gap between abstract low-fidelity concepts and the rich, experiential nature of immersive technologies.

As a result, we found that, to co-design with specialists outside the field of mixed reality and emerging technologies, it is necessary to create more realistic prototypes, even if they do not have full real-world functionality—which is also referred to as a "Wizard-of-Oz" approach [9]. Such prototypes lead to more realistic and feasible ideation, as opposed to ideas that are either too futuristic or too simplistic. Although this approach requires more time from the developers and researchers, it renders the sessions with specialists more productive. Moreover, due to the limited availability of specialists, it is justifiable for researchers to spend more time on prototypes to minimise wasted time and the number of sessions needed with domain experts.

**User privacy.** While one of the critical concerns in using technology in healthcare is the privacy of the patient's data—or, in the context of forensic medicine, the deceased person's data—the use of AI and immersive technologies has raised concerns about the privacy of system users. This concern is less relevant to current practice in forensic medicine, where data is primarily in written format or converted to text from voice recordings by human transcribers. Therefore, while systems like our automatic documentation may utilise voice or procedure recordings to increase reliability, transparency, and streamline the workflow, there is a risk of capturing information that is not necessary for the case but that reveals private information or conversations of the users. Consequently, while these technologies may enhance reliability and transparency, they also pose a risk of privacy breaches, which has received negative feedback from experts.

**Stakeholders impacted by the technology.** User studies are a common evaluation method used to assess human-in-the-loop systems, which typically define a user as the person interacting with the system. However, this focus may obscure other, potentially more important, aspects of system evaluation. In our expert-centred studies, we also noticed this limitation, which necessitates complementary studies to develop a broader understanding of a system's value.

For example, in our report generation study [16], the previously mentioned issue of "user privacy" was identified as a main drawback of the system by several participants. This concern has been observed in other fields as well, such as the use of body-cameras for police officers [10]. Our participants perceive this feature as potentially violating their privacy, introducing undue responsibilities, and shifting attention from outcomes to individual actions. While this perspective may be valid from the users' point of view, it is also necessary to consider the perspectives of who use the outputs derived from the system, rather than the system itself only. In summary, it is crucial to evaluate the balance between the drawbacks and advantages of a feature for different stakeholders, not necessarily only the users, as this may support a more informed and balanced conclusion.

This issue is particularly important in sensitive areas such as forensic investigation and healthcare. Further studies may need to reconsider existing policies to support users while also enhancing transparency. Such considerations could include clarifying access rights—who can access specific information—and determining what may be derived from this information as evidence.

### 3.6 Research Value

Conventional studies involving standard participant groups are often completed efficiently, with fewer complications. However, the opportunity to conduct research within a leading forensic institution offered substantial advantages throughout this journey and proved to be highly rewarding.

This environment enabled us to engage with state-of-the-art technologies and best practices, and to collaborate with experienced professionals over an extended period exceeding three years. Although the number of expert participants included in these studies was limited, peer reviewers recognised the significant value of the specialist insights obtained, which will likely prove highly valuable for subsequent researchers in this field. These also improved the visibility of the studies and attracted media attention (e.g., [1, 13, 2, 21, 11]).

## 4 CONCLUSION

In this paper, we reported on the main challenges associated with working with a small group of user-experts and the lessons learned from expert-centred studies conducted during a PhD study that explores immersive forensic investigation. In particular, we collaborated with a forensic medicine institution and police personnel, each of which required distinct ethics approval processes and followed different timeframes. In addition, we faced challenges in recruiting participants, and addressed these through offering authorship and adapting our research methods. Reflecting on the challenges and benefits of working with experts in this specific context may assist future researchers in planning ahead, enabling them to better anticipate both potential difficulties and the benefits inherent in these types of collaborations.

Future research can reflect on their own lessons and experiences from expert-centred studies, identifying similarities and differences across various applications and even locations. Additionally, researchers could consider alternative approaches, such as separating the design and development phases from evaluation, and delegating evaluation to internal researchers, which we have not explored.

## ACKNOWLEDGMENTS

We would like to thank our collaborators at the Victorian Institute of Forensic Medicine (VIFM) for their support and cooperation. This work was funded by VIFM and partially supported by the Knut and Alice Wallenberg Foundation (KAW 2019.0024) and the Marcus and Amalia Wallenberg Foundation (MAW 2023.0130).